\documentstyle[12pt]{article}

\def\hybrid{\topmargin -20pt	\oddsidemargin 0pt
	\headheight 0pt	\headsep 0pt
	\textwidth 6.25in	
	\textheight 9.5in	
	\marginparwidth .875in
	\parskip 5pt plus 1pt	\jot = 1.5ex}

\def\baselinestretch{1.2}

\catcode`\@=11

\def\marginnote#1{}
\newcount\hour
\newcount\minute
\newtoks\amorpm
\hour=\time\divide\hour by60
\minute=\time{\multiply\hour by60 \global\advance\minute by-\hour}
\edef\standardtime{{\ifnum\hour<12 \global\amorpm={am}%
	\else\global\amorpm={pm}\advance\hour by-12 \fi
	\ifnum\hour=0 \hour=12 \fi
	\number\hour:\ifnum\minute<10 0\fi\number\minute\the\amorpm}}
\edef\militarytime{\number\hour:\ifnum\minute<10 0\fi\number\minute}
\def\draftlabel#1{{\@bsphack\if@filesw {\let\thepage\relax
   \xdef\@gtempa{\write\@auxout{\string
      \newlabel{#1}{{\@currentlabel}{\thepage}}}}}\@gtempa
   \if@nobreak \ifvmode\nobreak\fi\fi\fi\@esphack}
	\gdef\@eqnlabel{#1}}
\def\@eqnlabel{}
\def\@vacuum{}
\def\draftmarginnote#1{\marginpar{\raggedright\scriptsize\tt#1}}

\def\draft{\oddsidemargin -.2truein
	\def\@oddfoot{\sl preliminary draft \hfil
	\rm\thepage\hfil\sl\today\quad\militarytime}
	\let\@evenfoot\@oddfoot	\overfullrule 3pt
	\let\label=\draftlabel
	\let\marginnote=\draftmarginnote
   \def\@eqnnum{(\theequation)\rlap{\kern\marginparsep\tt\@eqnlabel}%
\global\let\@eqnlabel\@vacuum}  }

\def\preprint{\twocolumn\sloppy\flushbottom\parindent 2em
	\leftmargini 2em\leftmarginv .5em\leftmarginvi .5em
	\oddsidemargin -.5in	\evensidemargin -.5in
	\columnsep .4in	\footheight 0pt
	\textwidth 10.in	\topmargin  -.4in
	\headheight 12pt \topskip .4in
	\textheight 6.9in \footskip 0pt
	\def\@oddhead{\thepage\hfil\addtocounter{page}{1}\thepage}
	\let\@evenhead\@oddhead	\def\@oddfoot{}	\def\@evenfoot{} }

\def\numberbysection{\@addtoreset{equation}{section}
	\def\theequation{\thesection.\arabic{equation}}}

\def\underline#1{\relax\ifmmode\@@underline#1\else
	$\@@underline{\hbox{#1}}$\relax\fi}

\def\titlepage{\@restonecolfalse\if@twocolumn\@restonecoltrue
\onecolumn
     \else \newpage \fi \thispagestyle{empty}\c@page\z@
	\def\thefootnote{\fnsymbol{footnote}} }

\def\endtitlepage{\if@restonecol\twocolumn \else \newpage \fi
	\def\thefootnote{\arabic{footnote}}
	\setcounter{footnote}{0}}  

\catcode`@=12
\relax

\def\pr{Phys. Rev. \/}
\def\np{Nucl. Phys. \/}
\def\pl{Phys. Lett. \/}
\def\figcap{\section*{Figure Captions\markboth
	{FIGURECAPTIONS}{FIGURECAPTIONS}}\list
	{Figure \arabic{enumi}:\hfill}{\settowidth\labelwidth{Figure
999:}
	\leftmargin\labelwidth
	\advance\leftmargin\labelsep\usecounter{enumi}}}
 \relax
\def\tablecap{\section*{Table Captions\markboth
	{TABLECAPTIONS}{TABLECAPTIONS}}\list
	{Table \arabic{enumi}:\hfill}{\settowidth\labelwidth{Table
999:}
	\leftmargin\labelwidth
	\advance\leftmargin\labelsep\usecounter{enumi}}}
 \relax
\def\reflist{\section*{References\markboth
	{REFLIST}{REFLIST}}\list
	{[\arabic{enumi}]\hfill}{\settowidth\labelwidth{[999]}
	\leftmargin\labelwidth
	\advance\leftmargin\labelsep\usecounter{enumi}}}
 \relax
\makeatletter
\newcounter{pubctr}
\def\publist{\@ifnextchar[{\@publist}{\@@publist}}
\def\@publist[#1]{\list
	{[\arabic{pubctr}]\hfill}{\settowidth\labelwidth{[999]}
	\leftmargin\labelwidth
	\advance\leftmargin\labelsep
	\@nmbrlisttrue\def\@listctr{pubctr}
	\setcounter{pubctr}{#1}\addtocounter{pubctr}{-1}}}
\def\@@publist{\list
	{[\arabic{pubctr}]\hfill}{\settowidth\labelwidth{[999]}
	\leftmargin\labelwidth
	\advance\leftmargin\labelsep
	\@nmbrlisttrue\def\@listctr{pubctr}}}
 \relax
\makeatother
\newskip\humongous \humongous=0pt plus 1000pt minus 1000pt

\newif\ifdtup

\relax
\hybrid

\def\thefootnote{\fnsymbol{footnote}}
\def\be{\begin{equation}}
\def\ee{\end{equation}}
\def\ba{\begin{eqnarray}}
\def\ea{\end{eqnarray}}

\def\p{\partial}

\def\t{\tau}
\def\tt{\bar\tau}
\def\R{{\cal{R}}}

\begin{document}
\renewcommand{\theequation}{\thesection.\arabic{equation}}
\newcommand{\beq}{\begin{equation}}
\newcommand{\eeq}[1]{\label{#1}\end{equation}}
\newcommand{\ber}{\begin{eqnarray}}
\newcommand{\eer}[1]{\label{#1}\end{eqnarray}}
\begin{titlepage}
\begin{center}

\hfill CERN-TH.7471/94\\
\hfill LPTENS-94/29\\
\hfill hep-th/9410212\\

\vskip .2in

{\large \bf Curved Four-Dimensional Spacetime as Infrared Regulator
in Superstring Theories}
\vskip .4in

{\bf Elias Kiritsis and Costas Kounnas\footnote{On leave from Ecole
Normale Sup\'erieure, 24 rue Lhomond, F-75231, Paris, Cedex 05,
FRANCE.}}\\
\vskip
 .3in

{\em Theory Division, CERN,\\ CH-1211,
Geneva 23, SWITZERLAND} \footnote{e-mail addresses:
KIRITSIS,KOUNNAS@NXTH04.CERN.CH}\\

\vskip .3in

\end{center}

\vskip .2in

\begin{center} {\bf ABSTRACT } \end{center}
\begin{quotation}\noindent
We construct a new class of exact and stable superstring solutions in
which our four-dimensional spacetime is taken to be curved . We
derive in this space the full one-loop partition function in the
presence of non-zero $\langle F^a_{\mu\nu}F_a^{\mu\nu}\rangle=F^2$
gauge background as well as in an $\langle
R_{\mu\nu\rho\sigma}R^{\mu\nu\rho\sigma}\rangle=\R^2$ gravitational
background and we show that  the non-zero curvature,
$Q^2=2/(k+2)$, of  the spacetime provides an infrared regulator for
all $\langle[F^a_{\mu\nu}]^n[R_{\mu\nu\rho\sigma}]^m\rangle$
correlation functions. The
string one-loop partition function $Z(F,\R, Q)$ can be exactly
computed,
and it is IR and UV finite.
For $Q$ small we have thus obtained an IR regularization,
consistent with spacetime
supersymmetry (when $F=0,\R=0$) and modular invariance. Thus, it can
be
used to determine, without any infrared ambiguities, the one-loop
string radiative corrections on gravitational, gauge or Yukawa
couplings necessary for the string superunification predictions at
low energies.
\end{quotation}
\vskip 1.0cm
\begin{center}
Contributed to the Proceedings of the Trieste Spring School and
Workshop, 1994.
\end{center}
\vskip 1.cm
CERN-TH.7471/94 \\
October 1994\\
\end{titlepage}
\vfill
\eject
\def\baselinestretch{1.2}
\baselineskip 16 pt
\noindent
\section{Introduction}
\setcounter{equation}{0}

The four-dimensional superstring solutions in a flat background
\cite{cand}-\cite{gepner}
 define at  low energy effective supergravity theories
\cite{effcl},\cite{moduli}.
 A class of them successfully extends
the validity of the standard model up to the string scale,
$M_{string}$.
 The first main property of superstrings is that they are
ultraviolet-finite theories (at least perturbatively). Their
second important
property is that  they unify gravity with all other interactions.
This unification does not include  only the gauge interactions, but
also the  Yukawa  ones as well as the interactions among the scalars.
This String Hyper Unification (SHU) happens  at  large energy scales
$E_t\sim {\cal O}(M_{string})\sim 10^{17}~$GeV. At this energy scale,
however,
the first excited string states become important and thus the whole
effective low energy field theory picture breaks
down\cite{n4kounnas,ki,worm,kktopol}. Indeed, the
effective field theory of strings is valid only for  $E_t \ll
M_{string}$ by means of
the ${\cal O}(E_t/M_{string})^2$ expansion. It is then necessary to
evolve the SHU predictions to a lower scale $M_U < M_{string}$ where
the  effective field theory picture makes sense. Then, at $M_U$, any
string solution provides  non-trivial relations between  the gauge
and
Yukawa couplings, which can be written as

\begin{equation}
\frac{k_i}{\alpha_i(M_U)}=\frac{k_j}{\alpha_j(M_U)}+\Delta_{ij}(M_U).
\label{shu} \end{equation}

The above relation looks very similar to the well-known unification
condition
in Supersymmetric Grand Unified Theories (SuSy-GUTs) where the
unification scale is about $M_U\sim 10^{16}~$GeV and
$\Delta_{ij}(M_U)=0$ in the ${\bar {DR}}$ renormalization scheme; in
SuSy-GUTs the normalization constants $k_i$ are fixed $only$ for the
gauge couplings ($k_1=k_2=k_3=1$, $k_{em}=\frac{3}{8}$), but there
are
no relations among gauge and Yukawa couplings at all. In string
effective theories, however, the normalization constants ($k_i$) are
known for both gauge and Yukawa interactions. Furthermore,
$\Delta_{ij}(M_U)$ are calculable $finite$ quantities for any
particular string solution. Thus, the predictability of a given
string solution is extended for all low energy coupling constants
${\alpha_i(M_Z)}$ once the string-induced corrections
$\Delta_{ij}(M_U)$ are determined.

 This determination  however, requests string computations which
 we did not know, up to now, how to perform in generality. It turns
out
that
$\Delta_{ij}(M_U)$ are non-trivial functions of the vacuum
expectation values  of  some gauge singlet fields  \cite{moduli},
$~~\langle T_A\rangle =t_A$, the so-called moduli (the moduli fields
are flat
directions at the string classical level and they remain flat in
string perturbation theory, in the exact supersymmetric limit) :
\begin{equation}
{\Delta_{ij} (M_U)= {E_{ij}}+ F_{ij} (t_A)}.
\label{delta} \end{equation}
Here $F_{ij}(t_A)$ are modular forms,  which depend on the particular
string solution.  Partial results for
$F_{ij}$ exist in the exact supersymmetric limit in many string
solutions based on orbifold
\cite{orbifold} and fermionic constructions \cite{abk4d}.
 The finite part $E_{ij}$
is a function of $M_U/M_{string}$ and, at the  present time, it is
only
 approximately estimated \cite{moduli}. As we will see later $E_{ij}$
are, in principle,   well defined
calculable quantities once we perform our calculations  at the string
level where all interactions including gravity are  consistently
defined. The full string corrections   to the coupling constant
unification, $\Delta_{ij}(M_U)$, as well as the string corrections
associated to the soft supersymmetry-breaking parameters
 \begin{equation}
m_0,~ m_{1/2},~ A,~ B~{\rm  and}~ \mu,~~{\rm  at}~~M_U,
\label{soft} \end{equation}
are of main importance, since they fix  the strength of the gauge and
Yukawa interactions, the full spectrum of the
supersymmetric particles as well as the Higgs and the top-quark
masses at the low energy range $M_Z\leq E_{t}\leq {\cal O}(1)$ TeV.

In the case where supersymmetry is broken \cite{gcsbr},\cite{ssbr}
only
semi-quantitative results can be obtained at present; a much more
detailed
study and understanding are  necessary which is related to the
structure of soft breaking terms after the assumed supersymmetry
breaking \cite{fkpz}.

The main obstruction in determining the exact form of the string
radiative corrections $\Delta_{ij}(M_U)$ is strongly related to the
infrared divergences of the $\langle [F^a_{\mu\nu}]^2\rangle $
two-point correlation function in superstring theory. In field
theory, we
can avoid this problem using off-shell calculations. In first
quantized string theory we cannot do that since we do not know how to
go off-shell.
Even in field theory there are problems in defining an infrared
regulator for chiral fermions especially in the presence of
spacetime supersymmetry.

In \cite{cw} it was suggested to use a specific spacetime with
negative curvature in order to achieve consistent regularization in
the infrared. The proposed  curved space however is not useful for
string applications since it does not correspond to an exact
super-string solution.

Recently, exact and stable superstring solutions
have been constructed  using special four-dimensional spaces as
superconformal building blocks with
${\hat c}=4$ and $N=4$ superconformal
symmetry \cite{n4kounnas}, \cite{worm}. The full spectrum of string
excitations for the superstring solutions based on those
four-dimensional subspaces, can be derived using the techniques
developed in
ref. \cite{worm}. The main characteristic property of these solutions
is the existence of a mass gap $\mu ^2=Q^2/4$, which is proportional
to the curvature of the non-trivial  four-dimensional spacetime.
Comparing the spectrum in  a flat background with that in curved
space we observe a shifting of all massless states by an amount
proportional to the spacetime curvature, $\Delta m^2=Q^2/4$. What is
also interesting is that the shifted spectrum in the curved space is
equal for bosons and fermions due to the existence of a new
space-time supersymmetry defined in the curved spacetime
\cite{n4kounnas} \cite{worm}. Therefore, our curved space time
infrared
regularization  (CSIR) is consistent with supersymmetry and can be
used either in field theory or string theory.

In section 2 we define the four-dimensional superconformal system and
give
the modular-invariant partition function for some symmetric orbifold
ground states of the string.
In section 3 we show that we can deform the theory  consistently, by
switching  on a non-zero gauge field strength background $\langle
F^a_{\mu\nu}F_a^{\mu\nu}\rangle =F^2$ or a gravitational one,
$\langle R_{\mu\nu\rho\sigma}R^{\mu\nu\rho\sigma}\rangle=\R^2$ and
obtain the exact
regularized partition function $Z(Q,F,\R)$.
Our method of constructing this effective action automatically
takes into
account the back-reaction of the other background fields; stated
otherwise, the perturbation that turns on the constant gauge field
strength or curvature background is an exact (1,1) integrable
perturbation. The
second derivative with respect to $r$ of our deformed partition
function
$\p^2 Z(Q,F,\R)/\p F^2$ for $F,\R=0$ defines without any infrared
ambiguities the
complete  string one-loop corrections to the gauge coupling
constants.
In the $Q\to 0$ limit we recover the known partial results
\cite{moduli}.

\section{Superstrings in Curved Space Time}

As usual in order to construct a four-dimensional superstring
solution
one must saturate the superconformal anomaly ${\hat c}=10$ combining
two sub-systems.

i) the four-dimensional space time superconformal system  with ${\hat
c}=4+\epsilon$ and

ii) a  six-dimensional compact space with ${\hat c}= 6-\epsilon$.

In all our constructions we impose $\epsilon=0$ so that the internal
compact space can be chosen to be in one-to-one correspondence with
any possible construction in four-dimensional flat space. The idea
here is to replace the four free space time supercoordinates ${\hat
c}=4$ with a non-trivial ${\hat c}=4$ Euclidean spacetime which
shares similar
superconformal properties, namely an $N=4$ superconformal symmetry.
In ref. \cite{kikoulu} a large class of such spaces is found.
Although
all these solutions are exact superconformal systems due to the $N=4$
symmetry, we will restrict in what follows to the four $N=4$
realizations of ref.\cite{kprn4} and \cite{n4kounnas} which are based
on
(gauged-) Wess-Zumino-Witten models mainly because we know well their
characters and thus can construct explicitly the one-loop
partition of the full string model.

As we already mentioned above, all four-dimensional superstring
solutions
are of the form $F^{(4)}\otimes K^{(6)}$, while the curved
four-dimensional superstring solutions replace the four-dimensional
flat
space time ($F^{(4)}$)
with one of the following possibilities \cite{n4kounnas}:

1) $W_{k}^{(4)} \equiv U(1)_{Q}\otimes SU(2)_{k_1}$

2) $C^{(4)}_k \equiv [SU(2)/U(1)]_{k}\otimes
U(1)_{R}\otimes U(1)_{Q}$

3) $\Delta^{(4)}_k(A) \equiv [SU(2)/U(1)]_k
\otimes [SL(2,R)/U(1)_A]_{k+4}$

4) $\Delta^{(4)}_k(V) \equiv [SU(2)/U(1)]_k
\otimes [SL(2,R)/U(1)_V]_{k+4}$

The background $Q$ in cases 1) and 2)  is related to the level $k$
due to the $N=4$ algebra, $Q=\sqrt{2/(k+2)}$ and guarantees that
${\hat c}=4$ for any value of $k$.

 In the limit of weak curvature (large $k$) the  $W_{k}^{(4)}$ space
can  be interpreted as a topologically non-trivial four-dimensional
manifold of the form $R \otimes S^3$. The underlying superconformal
field theory associated
to $W_{k}^{(4)}$ includes a supersymmetric $SU(2)_{k}$ WZW model
describing the three coordinates of $S^3$ as well as a non-compact
dimension with a background charge, describing the scale factor of
the sphere \cite{n4kounnas},\cite{worm}. Furthermore this space
admits two covariantly constant
spinors and, therefore, respects up to two space-time supersymmetries
consistently with the $N=4$ world-sheet
symmetry \cite{wormclas,n4kounnas,worm}.
The explicit representation of the desired $N=4$ algebra is derived
in \cite{kprn4} and \cite{n4kounnas}, while the target space
interpretation
as a four-dimensional semi-wormhole space is given in
\cite{wormclas}.

The space  $C^{(4)}_k$ is  factorized  in two 2-d subspaces; for
small
curvatures the first subspace is described by the $SU(2)/U(1)$ bell,
while the second subspace $U(1)_{R}\otimes U(1)_Q$ defines a
two-dimensional cylinder. On the other hand, the spaces $W_{k}^{(4)}$
 and $C^{(4)}_k$  are related
to each other by target space duality transformation and both share
the $N=4$ superconformal properties. The explicit
realization of the $C^{(4)}_k$ space is given in \cite{n4kounnas}.
{}From the conformal theory viewpoint $C^{(4)}_{k}$  is based on the
supersymmetric gauged WZW model
$C^{(4)}_k \equiv [SU(2)/U(1)]_{k}\otimes
U(1)_{R}\otimes U(1)_{Q}$ with a background charge
$Q=\sqrt{2/(k+2)}$ in  one of the two coordinate currents
$U(1)_{Q}$. The other free  coordinate $U(1)_{R}$ is compactified
on a torus with radius $R=\sqrt{k}$.

 The $\Delta^{(4)}_k(A,V)$ spaces \cite{n4kounnas},\cite{worm} are
also factorized in two 2-d subspaces; the first one is the
$[SU(2)/U(1)]$ bell, while the second one is described by either the
$[SL(2,R)/U(1)_A]$ cigar(axial gauging) or the $[SL(2,R)/U(1)_V]$
trumpet
(vector gauging). In the $\Delta^{(4)}_k(A,V)$ spaces the elementary
fields are the $[SU(2)/U(1)]_{k}$ (compact) parafermionic
currents  as well as  the $[SL(2,R)/U(1)]_{k'}$ non-compact
parafermionic currents. The level $k'=k+4$, so that the total central
charge ${\hat c}$ remains equal to 4 for any value of $k$.

\subsection{The $W_k^{(4)}\otimes K^{(6)}$ partition function}

The basic rules of construction in curved space time are similar to
that of the orbifold construction \cite{orbifold}, the free 2-d
fermionic constructions \cite{abk4d},  and the Gepner construction
\cite{gepner} where one combines in a modular-invariant way the
world-sheet degrees of freedom consistently with unitarity and
spin-statistics of  the string spectrum. We will
choose as a first example the derivation of the string
spectrum in  background $W^{(4)}_k\otimes K^{(6)}$, where $K^{(6)}$
six-dimensional space. For definiteness we choose this space to be
one of the symmetric orbifold model, used in $(2,2)$
compactifications.

Since the world-sheet fermions of the  $W^{(4)}_k$ superconformal
system are free and since the $K^{(6)}$ internal space is the same as
in the $F^{(4)}\otimes K^{(6)}$  superstring solutions, we can easily
obtain the
partition function of $W^{(4)}_k\otimes K^{(6)}$, for $k$ even, in
terms of that of
$F^{(4)}\otimes K^{(6)}$:
\begin{equation}
Z_{W}[Q,\tau, \bar{\tau}]=[\Gamma(SU(2)_k)(\tau, {\bar
\tau})]~Z^F[\tau, {\bar \tau}],
\label{zwf1}
\end{equation}
where $\Gamma (SU(2)_k)$ is nothing but the contribution to the
partition function of the  bosonic coordinates $X^{\mu}$ of the
curved background $W^{(4)}$ divided by the contribution of the four
free coordinates of the $F^{(4)}$ flat space,
\begin{equation}
\Gamma (SU(2)_k)={1\over 2}[({\rm Im}\tau)^{\frac{1}{2}}
\eta(\tau){\bar
\eta}({\bar \tau})]^{3}~~\sum_{a,b=0}^{1}Z^{SU(2)}[^{a}_{b}]
{}.
\label{zwf2}
\end{equation}
\be
Z^{SU(2)}[^a_b]=e^{-i\pi k ab/2}\sum_{l=0}^ke^{i\pi
bkl/2}\chi_l(\tau){\bar \chi}_{l+a(k-2l)}({\bar
\tau})
\ee
where $\chi_l(\tau)$ are the characters of  $SU(2)_k$ (see for
example \cite{su2ch}) and the integer $l$ is equal to twice  the
$SU(2)$ spin  $l=2j$.
It is necessary to use this orbifoldized version of $SU(2)_{k}$ comes
in order to project out negative norm states of the $N=4$
superconformal
representations, \cite{worm}.

To obtain the above formula we have used the continuous series of
unitary representations
of the Liouville characters \cite{worm} which are generated by the
lowest-weight  operators,
\begin{equation}
e^{\beta X_L}\ ;\quad \beta=-\frac{1}{2}Q +ip\ ,
\label{contchar}
\end{equation}
having positive conformal weights $h_p=Q^2/8+p^2/2$.
The fixed imaginary part in the momentum $iQ/2$ of the plane waves
is due to the non-trivial dilaton motion.

As a particular example we give below the partition function of the
$Z^2 \otimes Z^2$ symmetric orbifolds\cite{orbifold}, \cite{abk4d},
$W^{(4)}_k \otimes T^{(6)}$/$(Z^{2}\otimes Z^{2})$, for type-II and
heterotic constructions:
$$
Z^W_{II}[Q;\tau,{\bar \tau}]~=~ {\Gamma(SU(2)_{k})\over {\rm Im}\t
{}~\eta^2\bar\eta^2}\times
{1\over 16}\sum_{\alpha,\beta,{\bar \alpha},{\bar
\beta}=0}^{1}\sum_{h_1,g_1,h_2,g_2}
Z_1[^{h_1}_{g_1}]Z_2[^{h_2}_{g_2}]Z_3[^{-h_1-h_2}_{-g_1-g_2}]\times
$$
\be
(-)^{\alpha+\beta+\alpha\beta}
\frac{\vartheta[^{\alpha}_{\beta}]}{\eta}
\frac{\vartheta[^{\alpha+h_1}_{\beta+g_1}]}{\eta}
\frac{\vartheta[^{\alpha+h_2}_{\beta+g_2}]}{\eta}
\frac{\vartheta[^{\alpha-h_1-h_2}_{\beta-g_1-g_2}]}{\eta}
{}\times (-)^{{\bar\alpha}+{\bar\beta}+\bar\alpha\bar\beta}
\frac{{\bar\vartheta}[^{{\bar\alpha}}_{{\bar\beta}}]}
{{\bar\eta}}
\frac{{\bar\vartheta}[^{{\bar\alpha}+h_1}_{{\bar\beta}+g_1}]}
{{\bar\eta}}
\frac{{\bar\vartheta}[^{{\bar\alpha}+h_2}_{{\bar\beta}+g_2}]}
{{\bar\eta}}
\frac{{\bar\vartheta}[^{{\bar\alpha}-h_1-h_2}_{{\bar\beta}-g_1-g_2}]}
{{\bar\eta}}\label{type2}
\end{equation}
where $Z_i[^{h_i}_{g_i}]$ in (\ref{type2}) stands for the partition
function of two twisted bosons with twists ($h_i,g_i)$. The untwisted
part $Z_i[^{0}_{0}]$
is equal to the moduli-dependent two-dimensional lattice
$\Gamma(2,2)[T_i,U_i]$/$(\eta{\bar\eta})^2$.
The definition of the $\vartheta$-function we use is
\be
\vartheta[^{a}_{b}](v|\tau)=\sum_{n\in
Z}e^{i\pi\tau(n+a/2)^2+2i\pi(n+a/2)(v+b/2)}
\ee
In the heterotic case, a modular-invariant partition function  can be
easily obtained using the heterotic map \cite{llsmap},
\cite{gepner}. It consists in
replacing in (\ref{type2}) the $O(2)$ characters associated to the
right-moving fermionic coordinates ${\bar\Psi}^{\mu}$, with the
characters of either $O(10)\otimes E_8$:
\begin{equation}
(-)^{{\bar\alpha}+{\bar\beta}+\bar\alpha\bar\beta}
\frac{{\bar\vartheta}[^{\bar\alpha}_{\bar\beta}]}{{\bar\eta}}
\rightarrow
\frac{{\bar\vartheta}[^{\bar\alpha}_{\bar\beta}]^{5}}{{\bar\eta}^5}
{1\over
2}\sum_{\gamma,\delta}\frac{{\bar\vartheta
[^{\gamma}_{\delta}]}^{8}}{{\bar\eta}^8}\label{pfheta}
\end{equation}
or $O(26)$:
\begin{equation}
(-)^{{\bar\alpha}+{\bar\beta}+\bar\alpha\bar\beta}
\frac{{\bar\vartheta}[^{\bar\alpha}_{\bar\beta}]}{{\bar\eta}}
\rightarrow
\frac{{\bar\vartheta[^{\bar\alpha}_{\bar\beta}]}^{13}}
{{\bar\eta}^{13}}.
\label{pfhetb}
\end{equation}
Using the map above, the heterotic partition function  with
$E_{8}\otimes E_{6}$ unbroken gauge group is:

$$
Z^W_{het}[Q; \tau,\tt]={\Gamma(SU(2)_{k})\over {\rm Im}\t
{}~\eta^2\bar\eta^2}\times
{1\over 16}
\sum_{\alpha,\beta,{\bar\alpha},{\bar\beta}=0}^{1}
\sum_{h_1,g_1,h_2,g_2}
{Z_1[^{h_1}_{g_1}]Z_2[^{h_2}_{g_2}]Z_3[^{-h_1-h_2}_{-g_1-g_2}]}\times
$$

\be
(-)^{\alpha+\beta+\alpha\beta}
\frac{\vartheta[^{\alpha}_{\beta}]}{\eta}
\frac{\vartheta[^{\alpha+h_1}_{\beta+g_1}]}{\eta}
\frac{\vartheta[^{\alpha+h_2}_{\beta+g_2}]}{\eta}
\frac{\vartheta[^{\alpha-h_1-h_2}_{\beta-g_1-g_2}]}{\eta}\times
{1\over
2}\sum_{\gamma,\delta}\frac{{\bar\vartheta}
[^{\gamma}_{\delta}]^8}{{\bar\eta}^8}\frac{{\bar\vartheta}
[^{\bar\alpha}_{\bar\beta}]}{{\bar\eta}^5}
\frac{{\bar\vartheta}[^{{\bar\alpha}+h_1}_{{\bar\beta}+g_1}]}
{{\bar\eta}}
\frac{{\bar\vartheta}[^{{\bar\alpha}+h_2}_{{\bar\beta}+g_2}]}
{{\bar\eta}}
\frac{{\bar\vartheta}
[^{{\bar\alpha}-h_1-h_2}_{{\bar\beta}-g_1-g_2}]}
{{\bar\eta}}
\label{het}
\end{equation}

The mass spectrum of bosons and fermions in both the  heterotic and
type-II constructions is degenerate due to the existence of
space-time
supersymmetry defined in the $W^{(4)}_k$ background. The heterotic
constructions are  $N=1$ spacetime supersymmetric while in the
type-II construction one obtains  $N=2$ supersymmetric solutions.

The boson (or fermion) spectrum is obtained  by setting to $+1$ (or
to $-1$)
the statistical factor,
$(-)^{\alpha+\beta+{\bar\alpha}+{\bar\beta}+\alpha\beta+
\bar\alpha\bar\beta}$, in the type-II
construction, while one must set the statistical factor
$(-)^{\alpha+\beta}$=$+1$ (or $-1$) in the heterotic constructions.
In order to derive the lower-mass levels we need the  behaviour of
the bosonic and fermionic part of the partition function in the limit
where ${\rm Im}\tau$ is large (${\rm Im}\tau \rightarrow \infty$).
This behaviour can be easily derived from the above partition
functions.
\begin{equation}
Z^{W}(Q;\tau,{\bar\tau})\longrightarrow {\rm C}[{\rm
Im}\tau]^{-1}~e^{-\frac{{\rm Im}\tau}{2(k+2)}}.
\label{lomass}
\end{equation}
The above behaviour is universal and does not depend on the choice of
$K^{6}$ internal $N=(2,2)$ space. Only the multiplicity factor $C$
(positive for bosons and negative for fermions) depends on the
different constructions and it is always proportional to the number
of the lower-mass level states $\mu^2=1/[2(k+2)]=Q^2/4$. If we
replace the $W^{(4)}_k$ with any one of the other $N=4$ ${\hat c}=4$
spaces, $C^{(4)}_k$,$\Delta^{(4)}_k(A,V)$, we get identical infrared
mass shifting $\mu$.

As we will see in the next section,  the induced mass $\mu$ acts as a
well-defined  infrared regulator for all the on-shell correlation
functions and in particular for the two-point function correlator
$\langle F^{a}_{\mu\nu}F_{a}^{\mu\nu}\rangle$ (and  $\langle
R_{\mu\nu\rho\sigma}R^{\mu\nu\rho\sigma}\rangle$) on the torus, which
is associated
to the one-loop string corrections on the  gauge coupling constant.

\section{Non-zero ${\bf F^{a}_{\mu\nu}}$ and  $
R_{\mu\nu}^{\rho\sigma}$ Background in Superstrings}

Our aim is to define the  deformation of the two-dimensional
superconformal theory  which corresponds to a non-zero field strength
$F^{a}_{\mu\nu}$ background and find the integrated  one-loop
partition function ${\bf Z}^{W}(Q,F,\R)$,  where $F$ is by the
magnitude
of the field strength,
$F^2 \equiv \langle F^{a}_{\mu\nu}F_{a}^{\mu\nu}\rangle$ and $\R$ is
that of the curvature,  $\langle
R_{\mu\nu\rho\sigma}R^{\mu\nu\rho\sigma}\rangle=\R^2$.

\begin{equation}
{\bf Z}^{W} [Q,F,\R]=\frac{1}{V(W)} \int_{\cal F}
\frac{ d\tau d{\bar\tau} }{ ({\rm Im}\tau)^2 }
Z^{W}[Q,F,\R;\tau,{\bar\tau}]
\label{intpart}
\end{equation}
where $V(W)$ is the volume  of the $W^{(4)}_k$ space; modulo the
trivial infinity which corresponds to the one non-compact dimension,
the remaining three-dimensional compact space is that of the
three-dimensional sphere.
In our normalization:
$$
V(SU(2)_k)=\frac{1}{8\pi} (k+2)^{\frac{3}{2}}
$$
so that it matches in the flat limit with the conventional flat space
contribution.

 In flat space, a small non-zero  $F_{\mu\nu}^a$ background gives
rise
to an infinitesimal deformation  of the 2-d $\sigma$-model action
given by,
\begin{equation}
\Delta S^{2d}(F^{(4)})=\int dzd{\bar z}F_{\mu\nu}^a[x^{\mu}
\partial_z x^{\nu}+\psi^{\mu}\psi^{\nu}]{\bar J}_a
\label{fdef}
\end{equation}
Observe that for $F^a_{\mu\nu}$ constant (constant magnetic field),
the left moving operator $[x^{\mu} \partial_z
x^{\nu}+\psi^{\mu}\psi^{\nu}]$ is not a well-defined $(1,0)$ operator
on the world sheet. Even though  the right moving Kac-Moody current
${\bar J}_a$ is a well-defined $(0,1)$ operator, the total
deformation
is not integrable in flat space. Indeed, the 2-d $\sigma$-model
$\beta$-functions are not satisfied in the presence of a constant
magnetic field. This follows from the fact that there is a
$non$-$trivial$ $back$-$reaction$ on the gravitational background due
the non-zero
magnetic field.

The important property of $W^{(4)}_k$ space is that we can solve this
back-reaction ambiguity. First observe that the deformation that
corresponds to a constant magnetic field
$B_i^a=\epsilon_{oijk}F_a^{ik}$ is a well-defined
(1,1) integrable deformation, which breaks the $(2,2)$ superconformal
invariance but keeps the $(1,0)$:
\begin{equation}
\Delta S^{2d}(W^{(4)}_k)=\int dzd{\bar
z}B^a_i[I^i+\frac{1}{2}\epsilon^{ijk}\psi_{j}\psi_{k}]{\bar J}_a
\label{fdef1}
\end{equation}
where $I^i$ is anyone of the $SU(2)_{k}$ currents.
The deformed partition function is not zero due to the breaking of
$(2,2)$ supersymmetry.
In order to see that this is the correct replacement of the Lorentz
current in the flat case, we will write the SU(2) group element as
$g=\exp[i{\vec\sigma}\cdot{\vec x}/2]$ in which case
$I^{i}=kTr[\sigma^{i}
g^{-1}\p g]=ik(\p x^{i}+\epsilon^{ijk}x_j\p x_k+{\cal{O}}(|x|^3))$.
In the flat limit the first term corresponds to a constant gauge
field
and thus pure gauge so the only relevant term is the second one that
corresponds to constant magnetic field in flat space.
The $\cal{R}$ perturbation is
\be
\Delta S({\cal{R}})=\int dzd\bar
z{\cal{R}}\left[I^{3}+\psi^{1}\psi^{2}\right]
\bar I^{3}
\ee
In $\sigma$-model language, in the flat limit it gives a metric
perturbation
\be
\delta (ds^2)=-{\cal{R}}\left[x^{1}dx^{2}-x^2dx^{1}\right]^2
\ee
with constant Riemann tensor and scalar curvature equal to
$6{\cal{R}}$.
There is also a  non-zero antisymetric tensor with
$H_{123}=2\sqrt{{\cal{R}}}$
and dilaton $\delta
\Phi={\cal{R}}\left[(x^1)^2+(x^2)^2+4(x^3)^2\right]/4$.

Due to the rotation invariance in $S^3$ we can choose
$B_i^a=F\delta_i^3$ without loss of generality . The vector $r_a$
indicates the
direction in the gauge group space of the right-moving affine
currents.
Looking at the $\sigma$-model representation of this perturbation,
we can observe that the $F_{\mu\nu}$ of this background gauge field
is a monopole-like gauge field on $S^3$ and its lift to the tangent
space is constant. Thus at the flat limit of the sphere it goes to
the constant $F_{\mu\nu}$ background of
flat space.

The moduli space of the $F,\R$ deformation is then given by the
$SO(1,n)/SO(n)$ Lorentzian-lattice boostings with $n$ being  the rank
of the right-moving gauge group. We therefore conclude that the
desired partition function ${\bf Z}^{W} (Q,F,\R)$  is given in terms
of the moduli of the $\Gamma(1,n)$ lorentzian lattice. The constant
gravitational background
$R^{ij}_{kl}=\R\epsilon^{3ij}\epsilon_{3kl}$ can also be
included exactly
by an extra boost, in which case the lattice becomes
$\Gamma(1,n+1)$.

Let us denote by  $Q$ the fermionic lattice momenta associated to
the left-moving $U(1)$ current $\partial H=\psi^1\psi^2$, by $I$
the charge lattice of the left-moving $U(1)$ current associated to
the $I_3$ current of $SU(2)_k$,
by $\bar Q$ the charge lattice of a right U(1) which is part of
the Cartan algebra of the non-abelian right gauge group and
by $\bar I$ the charge lattice of the right-moving $U(1)$ current
associated
to the $\bar I_3$ current of $SU(2)_k$.
In terms of these charges the part of the undeformed partition
function
that depends on them can be written as
\be
Tr[\exp[-2\pi \rm{Im}\tau (L_{0}+\bar L_{0})
+2\pi i\rm{Re}\tau (L_{0}-\bar L_{0})]]
\end{equation}
where
\be
L_{0}={1\over 2}Q^2+{I^2\over k}\;,\;\bar L_{0}={1\over 2}\bar
Q^2+{\bar I^2\over k}
\ee
The (1,1) perturbation that turns on a constant gauge field strength
$F$ as well
as a constant curvature $\R$ background
produces an special 2-parameter $O(2,2)$ boost in the charge lattice
above,
which transforms $L_{0}$ and $\bar L_{0}$ to
\be
L_{0}'=L_{0}+{\cosh\psi-1\over 2}\left( {(Q+I)^2\over
k+2}+\left(\cos\theta{\bar I\over \sqrt{k}}
+\sin\theta {\bar Q\over \sqrt{2}}\right)^2\right)
+
\ee
$$+\sinh\psi{(Q+I)\over \sqrt{k+2}}\left(\cos\theta{\bar I\over
\sqrt{k}}
+\sin\theta{ \bar Q\over \sqrt{2}}\right)
$$
and
\be
L'_{0}-\bar L'_{0}=L_{0}-\bar L_{0}
\ee

The parameters $\theta$  and $\psi$ are related to the constant
background fields
$F$ and $\R$ by\footnote{The $k$-dependence is such that there is
smooth flat
space limit.}
\be
F={\sinh\psi\sin\theta\over\sqrt{2(k+2)}}\;\;,\;\;
\R={\sinh\psi\cos\theta\over
\sqrt{k(k+2)}}
\ee
so that
\be
L_{0}'-L_{0}=(Q+I)\left(\R\bar I
+F{ \bar Q}\right)+
\ee
$$+
{\sqrt{1+(k+2)(2F^2+k\R^2)}-1\over 2}\left({(Q+I)^2\over
k+2}+{\left(\R\bar I+
+F\bar Q\right)^2\over (2F^2+k\R^2)}\right)
$$

The first term is the standard perturbation while the second term is
the back-reaction necessary for conformal and modular invariance.
Expanding the partition function in a power series in $F,\R$
\begin{equation}
Z^{W}(Q,F,\R)=\sum_{n,m=0}^{\infty}{F^{n}\R^{m}\over
n!m!}Z_{n,m}^{W}(Q)
\end{equation}
we can obtain the integrated correlators $\langle F^n R^m\rangle$.
For $\langle F^2\rangle$, $\langle FR\rangle$ and $\langle
R^2\rangle$ we
obtain:
\be
Z_{2,0}^{W}(Q)=-{2\pi{\rm Im}\t}\left[2(Q+I)^2+(k+2)\bar Q^2-
8\pi{\rm Im}\t(Q+I)^2\bar Q^2\right]
\ee
\be
Z_{1,1}^{W}(Q)=-{2\pi{\rm Im}\t}\left[k+2-{8\pi\rm{Im}
\t}{(Q+I)^2}\right]\bar Q\bar I
\ee
\be
Z_{0,2}^{W}(Q)=-{2\pi\rm{Im}\t}\left[k(Q+I)^2+(k+2)\bar I^2
-8\pi\rm{Im}\t(Q+I)^2\bar I^2\right]
\ee

The charges $Q_{i}$ in the above formula act in the respective
$\vartheta\left[^{\alpha}_{\beta}\right](\tau,v)$-functions as
differentiation with respect to $v$.
In particular $Q$ acts in the $\vartheta [^{\alpha}_{\beta}]$ in
equ. (7),
$I,\bar I$ act in the level-$k$ $\vartheta$-function present in
$\Gamma(SU(2)_{k})$ (due to the parafermionic decomposition), and
$\bar Q$ acts on one of  the right $\bar \vartheta$-functions.

We are interested in the one-loop correction to the
gauge
couplings, which is proportional to $Z_{2,0}^{W}(Q)$.
We can use the Riemann identity to transform the sum over  the
($\alpha,\beta)$ $\vartheta$-function characteristics (with non-zero
$v$)
that appear in (7,10) into
\be
{1\over 2}\sum_{a,b=0}^{1}(-)^{\alpha+\beta+\alpha\beta}
\vartheta[^{\alpha}_{\beta}](v|\tau)
\vartheta[^{\alpha+h_1}_{\beta+g_1}](0|\t)
\vartheta[^{\alpha+h_2}_{\beta+g_2}](0|\t)
\vartheta[^{\alpha-h_1-h_2}_{\beta-g_1-g_2}](0|\t)
=\ee
$$
=\vartheta[^1_1](v/2|\t)\vartheta[^{1-h_1}_{1-g_{1}}](v/2|\t)
\vartheta[^{1-h_2}_{1-g_{2}}](v/2|\t)
\vartheta[^{1+h_1+h_2}_{1+g_{1}+g_{2}}](v/2|\t)
$$
In this representation the charge operators are derivatives with
respect to $v$.

We will focus for simplicity to heterotic $Z_{2}\times Z_{2}$
orbifolds.
In this case all the characteristics in eq. (\ref{het}) take the
values
$0,1$.
The only non-zero contribution appears when one of the pairs
$(h_i,g_i)$ of
twists  is $(0,0)$ and the rest non-zero. There are three sectors
where
two out of the four fermion $\vartheta$-functions depend only on
$v/2$; they give non-zero contribution only when both derivatives
with respect to $v$  act on them. We have in total three $N=2$
sectors; the $N=4$ and the  $N=1$ sectors give zero contribution in
$Z_{2,0}(Q)$ for the  $Z^{2}\otimes Z^{2}$ orbifold model.
For other orbifold models there will be non-zero contributions from
the N=1 sectors.
Using  the fact that the contribution to the partition function of
the twisted bosons cancels (up to a constant that is proportional to
the number of fixed points)  that of the twisted fermions, and  also
the identity $\vartheta'(0)/2\pi=\eta^3$, we obtain the following
formula for $Z_{2,0}(Q)$:

\be
Z^{A}_{2,0}(Q)=- \sum_{i=1}^{3} \int_{\cal F} {d\t
d\tt\over {\rm Im}\t}{\Gamma(SU(2))\over
V(SU(2))}{\Gamma^{i}_{2,2}(T_{i},U_{i})\over \bar \eta^{24}}
\left[\bar Q_{A}^2-{1\over 4\pi{\rm Im}
\t}\right]2\bar\Omega(\tt)\label{univ}
\ee
where $A$ indicates the appropriate gauge group ($E_{8}$,$E_{6}$ or
$U(1)$),
$\bar Q_{A}$ is the associated charge operator, normalized so that it
acts as
${i\over \pi}{\p\over \p\tt}$ on the $\vartheta$-functions and
$\bar\Omega=\bar\Omega_8\bar\Omega_6$
with
\def\vt{\bar\vartheta}
\be
\bar\Omega_8(\tt)={1\over
2}\sum_{a,b=0}^{1}\bar\vartheta^8[^{a}_{b}]\;\;,\;\;
\bar\Omega_6(\tt)={1\over 4}\left[
\vt^8_{2}(\vt^4_3+\vt_{4}^4)-\vt_4^8(\vt_3^4+\vt_2^4)+
\vt_3^8(\vt_2^4-\vt_4^4)\right]
\ee
Thus the one-loop corrected gauge coupling constant can be written as
\be
{16\pi^2\over g_{A}^2(Q)}={16\pi^2\over g_{A}^2(M_{str})}+Z^{A}_{2,0}
\ee
Eq. (\ref{univ}) applies to any 4-d symmetric orbifold string model,
the only things that change are the moduli contribution $\Gamma^i$
and the specific form of $\bar\Omega$.
This formula differs from that of \cite{moduli} since it includes the
so-called universal contribution and is UV and IR $finite$.
In particular the back-reaction of gravity is included exactly and
contributes to the universal terms.
Taking differences between different gauge groups we obtain the
regularized
form of the result of \cite{moduli}.
The only difference from their formula is the replacement of the flat
space
contribution by $\Gamma(SU(2))/V(SU(2))$.
Our result is $explicitly$ $modular$ $invariant$ and $finite$.

In order to clearly see  how  the $W^{(4)}_{k}$ acts as an IR
regulator, it is convenient  to perform the summation on the spin
index $l$ of the $SU(2)$ characters. This sum can be done
analytically and one obtains the following surprising (and eventually
useful)  identity
\be
\Gamma(SU(2)_k)=\sqrt{\rm{Im}\t}{(k+2)^{3/2}\over 4\pi}\left[{\p
Z(R)\over \p R}|_{R^2=k+2}-{1\over 2}{\p Z(R)\over \p
R}|_{R^2=(k+2)/4}\right]
\ee
where $Z(R)$ is the $\Gamma (1,1)$ lattice contribution of the torus:
\be
Z(R)= \sum_{m,n}\exp\left[ {i\pi\t\over 2}({m \over
R}+nR)^2-{i\pi\tt\over 2} ({m \over R}-nR)^2\right]
\ee
Using the identity  above, $Z_{2}(Q)$ becomes,
\be
Z^A_{2}(Q)=  \sum_{i=1}^{3} 2 \int_{\cal{F}}{d\t d\tt\over {\rm
Im}\t^2}
{\rm Im}\t^{1\over 2}\left[Z'(R)|_{k+2}-{1\over
2}Z'(R)|_{(k+2)/4}\right]
\left[{\rm Im}\t
\Gamma^i_{(2,2)}(T_{i},U_{i})\Sigma^{A}\right]\label{32}
\ee
The function $\Sigma^A$ depends on the gauge group in question
and its constant part is $C(g_{A})-T(R^{i}_{A})$.
For example, in the $E_8$ case it is given by
\be
\Sigma^{E_8}=-2{\bar\Omega_6\over \bar\eta^{24}}\left[{i\p\over
\pi\p\tt}-{1\over 4\pi\rm{Im}\t}\right]\bar\Omega_8
\ee
The combination ${i\p\over \pi\p\tt}-{1\over 4\pi\rm{Im}\t}$ is a
covariant derivative on modular forms.

This is the final form for the complete string one-loop radiative
correction
to the appropriate gauge couplings.
This result is finite and manifestly invariant under the target space
duality group that acts on the $T_{i},U_{i}$ moduli.
We see in particular that the (regulated) integrand in our case is
related
to the partition function of a (3,3) lattice at special values of the
(3,3) moduli.
The derivative with respect to the $R$ modulus is responsible for the
regulation of the IR.
In order to see this  we will evaluate the part of the
radiative correction coming from the low-lying states, which in the
unregulated case is responsible for the IR divergence.
This is achieved by replacing the (2,2) lattice contribution in eq.
(\ref{32}) by 1 and leaving apart for the moment the universal
contribution:
\be
Z^{m=\mu}_{2}(Q)=\left[\sum_{i=1}^{3}b_i\right] \int_{\cal{F}}{d\t
d\tt\over
{\rm Im}\t} {\rm Im}\t^{1 \over 2}\left[Z'(R)|_{k+2}-{1\over
2}Z'(R)|_{(k+2)/4}\right]
\ee
As expected, $Z^{m=\mu}_{2}(Q)$ turns out to be finite and for large
$R$ behaves like
\be
Z^{m=\mu}_{2}(Q)=
(b_1+b_2+b_3)[\log(M_{str}^{2}/Q^2) +2c_0]+...\label{c}
\ee
where the dots stand for terms vanishing in the limit $Q\to 0$.
We define $M_{str}^2$ to be the mass of the lowest lying oscilator
state in the string spectrum ($M_{str}^2=1/\alpha'$).
The constant $c_{0}$ can be computed exactly with the result
\be
c_{0}={3\over 2}-{1\over 2}\log(\pi/2)-{1\over 2}\psi(1)
-{3\over 4}\log(3)=0.738857...
\ee
Observe that the coefficient $b_{1}+b_{2}+b_{3}=3C(g_{a})-T(R_{a})$
is nothing but the
$N=1$ $\beta$-function coefficient.
The constant coefficient $c_{0}$, together with that of massive
states
$F(T_{i},U_{i})$ as well as the universal contribution
define unambiguously the string scheme and can thus be compared with
the field theory result (regularized in the IR in the same way as
above) in any UV scheme, for instance the conventional $\bar {DR}$.
Although this coefficient is small, one has to compute the parts left
over
including the moduli dependence.
In particular the universal contribution can be important.
We calculate here the universal contribution due to would be massless
states, e.g. the constant part of $\bar\Omega/\bar\eta^{24}$.
This is equal to
\be
{60\over \pi}\int_{\cal{F}}{d^2\t\over
{\rm{Im}\t^2}}\sqrt{\rm{Im}\t}\left[Z'(R)|_{k+2}-{1\over
2}Z'(R)|_{(k+2)/4}\right]=20+{\cal{O}}(1/k)
\ee
This contribute to the coefficient $c_{0}$ in (\ref{c}) equal to
$1/3$ for $E_{8}$ and $-5/21$ for $E_{6}$.
This implies that a full calculation is necessary, namely the
contributions from all massive states, in order to find the exact
string scheme.
The explicit calculation of $Z_{2}(Q)$ at one-loop, including the
moduli dependence, is under way \cite{kkk}.

\section{Conclusions}

We have presented an IR regularization for string theory (and field
theory) induced by the curvature of spacetime as well as by
non-trivial  dilaton and axion fields.
This regularization preserves a form of spacetime supersymmetry and
gives masses to all massless fields (including chiral fermions) that
are proportional to the curvature.

In the regulated string theory we can compute exactly the one-loop
effective action for arbitrarily large, constant, non-abelian
magnetic
fields.
Using this result among other things, we can compute unambiguously
the
string-induced one-loop threshold corrections to the gauge couplings
as functions of the moduli. The eventual integral to be done contains
a special subclass of sums associated with (3,3) lattices.

\vskip 1cm

\centerline{\bf Acknowledgements}

We would like to thank the organizers of the
Spring School and Workshop 94 in ICTP, Trieste, for their warm
hospitality.\\
One of us (C.K.) was  supported in part by EEC contracts
SC1$^*$-0394C and SC1$^*$-CT92-0789.

\end{document}